\documentclass[english,english,aps,prl,floats,showpacs,twocolumn]{revtex4-1}
\usepackage[T1]{fontenc}
\usepackage[latin9]{inputenc}
\usepackage{color}
\usepackage{babel}
\usepackage{amssymb}
\usepackage{graphicx}
\usepackage{esint}
\usepackage[unicode=true,
 bookmarks=true,bookmarksnumbered=false,bookmarksopen=false,
 breaklinks=false,pdfborder={0 0 1},backref=false,colorlinks=true]
 {hyperref}
\hypersetup{pdftitle={Dipolar ordering and domains in crystals of Mn12 Ac molecular magnets},
 pdfauthor={D. A. Garanin}}
\begin{document}

\title{Dipolar ordering and domains in crystals of Mn$_{12}$ Ac molecular
magnets}

\author{D. A. Garanin}

\affiliation{Physics Department, Lehman College and Graduate School, The City
University of New York, 250 Bedford Park Boulevard West, Bronx, NY
10468-1589, U.S.A.}

\date{\today}
\begin{abstract}
Dipolar coupling in crystals of the Mn$_{12}$Ac molecular magnet
elongated along the anisotropy axis favors ferromagnetic ordering
below the mean-field Curie temperature 0.71 K. With the help of Monte
Carlo on crystals of up to one million Mn$_{12}$ molecules, it is
shown that ordering occurs at 0.36 K. The resulting state is split
into ferromagnetic domains with domain walls preferring the diagonal
orientation. Domain walls are pinned by the lattice at low temperatures.
Making the crystal shorter makes domains finer and smoothens out the
singularity at the transition, decreasing susceptibility in the domain
state. Hysteresis loops look ferromagnetic for prolate crystals and
antiferromagnetic for oblate crystals.
\end{abstract}

\maketitle

\section{Introduction}

Magnetic ordering in crystals of molecular magnets, caused by the
dipole-dipole interaction (DDI), is especially interesting because
of its close relation to the spectacular phenomenon of spin tunneling
under the barrier created by the strong uniaxial anisotropy \cite{frisartejzio96prl}
(for a recent review, see Ref. \cite{MM14springer}). The most famous
of molecular magnets is Mn$_{12}$Ac, each molecule of which behaves
as a giant spin $S=10$ that can point in two directions along the
easy axis. Mn$_{12}$Ac crystallizes into a body-centered tetragonal
lattice with the periods $a=b=17.319$ Å and $c=12.388$ Å, $c$ being
the direction of the magnetic easy axis $z$.

There is no exchange interaction between the molecules, the strongest
interaction being the DDI. In prolate crystals the DDI is creating
the dipolar field $B_{z}^{(D)}=0.0526$ T (Ref. \cite{garchu08prb})
that was also measured in experiment \cite{mchughetal09prb}. Such
strong dipolar bias affects the resonance condition for spin tunneling
and makes it collective. In addition to experimentally observed magnetic
deflagration \cite{suzetal05prl,garchu07prb}, fast dipolar-controlled
fronts of tunneling were proposed \cite{gar13prb}.

The possibility of dipolar ordering in different molecular magnets
was considered in Ref. \cite{feralo00prb}, where Monte Carlo simulations
on small systems were done. Energies of different types of dipolar
ordering in another molecular magnet Fe$_{8}$ were computed in Ref.
\cite{marchuaha01epl}. Neutron-diffraction experiments on Mn$_{12}$Ac
in transverse magnetic field \cite{luietal05prl} suggested ferromagnetic
ordering below about 1 K. Dipolar ordering was observed by magnetic
measurements in Fe$_{17}$ compound with $S=35/2$ \cite{evaetal06prl}
and in Mn$_{6}$ compound with $S=12$ \cite{moretal06prb}, both
materials having a small anisotropy.

The mean-field-based theory of dipolar ordering, domain-wall structure
and domain-wall motion in Mn$_{12}$Ac was developed in Ref. \cite{garchu08prb}.
Measurements of the magnetic susceptibility of Mn$_{12}$Ac on the
high-temperature side suggest its divergence at a finite temperature,
indicative of a phase transition \cite{milkensaryes10prb,bowenetal10prb}.
Theoretically, for elongated crystals with the aspect ratio exceeding
6 (Ref. \cite{gar10prbrc}) uniform ferromagnetic ordering has lower
energy than ordering with alternating ferromagnetic columns or planes
\cite{garchu08prb}. However, in Ref. \cite{gar10prbrc} by solving
the space-resolved Curie-Weiss equation it was shown that ordering
occurs into the ferromagnetic state with domains rather than into
the uniform ferromagnetic state. In Ref. \cite{burzurietal11prl}
magnetic ordering in Fe$_{8}$ was observed at $T=0.34$ K by magnetic
measurements.

The purpose of the work presented in this paper was to elicudate dipolar-driven
magnetic ordering in Mn$_{12}$Ac by large-scale Monte Carlo simulations.
Of a particular interest is dependence of the domain structure on
the aspect ratio of the crystals. Simulation of a ferromagnetic state
with domains requires a large crystal size, thus crystals containing
up to a million of magnetic molecules were used in computations. For
so large system sizes the long-range character of the DDI creates
a computational problem. Thus a modification of the Monte Carlo procedure
regarding updating the dipolar field was done that allowed a considerable
speed-up.

The main part of this article is organized as follows. The model is
introduced in Sec. \ref{sec:The-model}, where a short survey of the
previously obtained results is done. Sec. \ref{sec:Columns-and-domains}
is devoted to elucidation the role of coulumns of spins along the
$c$ axis, the main boulding block of the order in Mn$_{12}$Ac. Here
the optimal configuration of domain walls is found and estimations
for the domain structure are done. In Sec. \ref{sec:Numerical-method}
the numerical method, especially concerning the dipolar-field update,
is described in detail. Sec. \ref{sec:Numerical-results} presents
numerical results and compares them with predictions of Sec. \ref{sec:Columns-and-domains}.

\section{The model\label{sec:The-model}}

The model of the Mn$_{12}$Ac crystal includes pseudospin variables
$\sigma_{i}=\pm1$ for molecules at each lattice site $i$ of a boby-centered
tetragonal lattice. The magnetic moment of a molecule is $Sg\mu_{B}$
with $g=2.$ The dipole-dipole interaction in molecular magnets is
due to the longitudinal components of the spins only and can be written
in the form
\begin{equation}
\mathcal{H}\equiv-\frac{E_{D}}{2}\sum_{i,j}\phi_{ij}\sigma_{iz}\sigma_{jz},\label{Ham-DDI}
\end{equation}
 where $E_{D}=\left(Sg\mu_{B}\right)^{2}/v_{0}$ is the dipolar energy,
$E_{D}/k_{B}=0.0671$ K for Mn$_{12}$Ac, $v_{0}=abc$ is the unit-cell
volume and
\begin{equation}
\phi_{ij}=v_{0}\frac{3\left(\mathbf{e}_{z}\cdot\mathbf{n}_{ij}\right)^{2}-1}{r_{ij}^{3}},\qquad\mathbf{n}_{ij}\equiv\frac{\mathbf{r}_{ij}}{r_{ij}}.
\end{equation}
The position vector on one of the sublattices can be written as
\begin{equation}
\mathbf{r}_{i}=i_{a}a\mathbf{e}_{a}+i_{b}b\mathbf{e}_{b}+i_{c}c\mathbf{e}_{c},\qquad i_{a},i_{b},i_{c}=0,\pm1,\pm2,\ldots\label{riDef}
\end{equation}
and that on the other sublattice is given by the same expression with
the indices $i_{a},i_{b},i_{c}$ shifted by 1/2. Thus the reduced
DDI can be put into the form $\phi_{ij}=\phi\left(i_{a}-j_{a},i_{b}-j_{b},i_{c}-j_{c}\right)$
with
\begin{equation}
\phi\left(n_{a},n_{b},n_{c}\right)=\eta\frac{2\eta^{2}n_{c}^{2}-n_{a}^{2}-n_{b}^{2}}{\left(n_{a}^{2}+n_{b}^{2}+\eta^{2}n_{c}^{2}\right)^{5/2}},\qquad\eta\equiv\frac{c}{a},\label{phinnnDef}
\end{equation}
where $\eta=0.7153$ for Mn$_{12}$Ac.

The dipolar field on magnetic molecule $i$ is the sum over positions
of all other molecules $j$
\begin{equation}
B_{i,z}^{(D)}=\frac{Sg\mu_{B}}{v_{0}}D_{i,zz},\qquad D_{i,zz}\equiv\sum_{j}\phi_{ij}\sigma_{jz},\label{BviaD}
\end{equation}
where $D_{zz}$ is the reduced dipolar field. Inside a uniformly magnetized
ellipsoid with the symmetry axis $z$, the dipolar field is uniform,
so that one has $D_{zz}=\bar{D}_{zz}\sigma_{z}.$ Here
\begin{equation}
\bar{D}_{zz}=\bar{D}_{zz}^{(\mathrm{sph})}+4\pi\nu\left(1/3-n^{(z)}\right),\label{Dzzbar}
\end{equation}
$\nu$ is the number of magnetic molecules per unit cell ($\nu=2$
for Mn$_{12}$ Ac) and $n^{(z)}=0,$ $1/3,$ and 1 for a cylinder,
sphere, and disc, respectively. The reduced dipolar field in a sphere
$\bar{D}_{zz}^{(\mathrm{sph})}$ depends on the lattice structure.
For Mn$_{12}$Ac direct lattice summation yields $\bar{D}_{zz}^{(\mathrm{sph})}=2.155$
that results in $\bar{D}_{zz}^{(\mathrm{cyl})}=10.53$ for a cylinder
\cite{garchu08prb}. Then Eq.\ (\ref{BviaD}) yields the dipolar
field $B_{z}^{(D)}\simeq52.6$ mT in an elongated sample that was
also obtained experimentally \cite{mchughetal09prb}.

The ground-state energy in the above uniform state is given by
\begin{equation}
E_{0}=-(1/2)\bar{D}_{zz}E_{D},\qquad E_{D}\equiv\left(Sg\mu_{B}\right)^{2}/v_{0}.\label{E0Def}
\end{equation}
 The mean-field Curie temperature is given by \cite{garchu08prb}
\begin{equation}
T_{C}=E_{D}\bar{D}_{zz}/k_{B}\label{TCsmallDelta}
\end{equation}
that yields $T_{C}=0.707$ K.

States with ferromagnetically ordered planes alternating in the $a$
or $b$ directions in \emph{each} sublattice of Mn$_{12}$ Ac have
$\bar{D}_{zz}=9.480,$ independently of the sample shape \cite{garchu08prb}.
The state with alternating chains in each sublattice, directed along
the $c$ direction has a very close value $\bar{D}_{zz}=9.475.$ For
the two-sublattice antiferromagnetic ordering one obtains $\bar{D}_{zz}=8.102.$
Thus, in a strongly prolate ellipsoid of Mn$_{12}$ Ac ferromagnetic
ordering is preferred. Below it will become clear that for box-shape
crystals the results do not depend on the aspect ratio and the lowest-energy
state is ferromagnetic with domains.

\section{Columns and domains\label{sec:Columns-and-domains}}

One can calculate the reduced dipolar field produced by an infinitely
long column of ordered spins on one of its own sites. One obtains
$D_{zz}=\bar{D}_{zz}\sigma_{z}$, where
\begin{equation}
\bar{D}_{zz}(0,0)=2\sum_{n=1}^{\infty}\phi\left(0,0,n\right)=\frac{4}{\eta^{2}}\sum_{n=1}^{\infty}\frac{1}{n^{3}}=\frac{4\zeta(3)}{\eta^{2}}\label{DzzCol}
\end{equation}
 that for Mn$_{12}$Ac yields $\bar{D}_{zz}(0,0)=9.39742.$ This field
is much larger than the field produced by such a column at any site
that does not belong to it (see below). For this reason columns aligned
along the $c$ axis are very stable and can be considered as a building
block for magnetic ordering in Mn$_{12}$Ac.

Within the continuous approximation, the dipolar field produced by
an infinite column of ordered spins disappears outside the column.
In fact, however, some dipolar field due to the discreteness of the
lattice remains, $D_{zz}<0$ at the sites of the same sublattice and
$D_{zz}>0$ at the sites of the other sublattice {[}Eq. (3) of Ref.
\cite{feralo00prb}{]}. This field decreases extremely fast with the
distance from the column, so that it is sufficient to take into account
only the interaction with nearest neighbors. With distances specified
by $n_{\rho}$ and $n_{c}$ for Mn$_{12}$Ac one has
\begin{eqnarray}
\bar{D}_{zz}\left(\frac{1}{\sqrt{2}},\frac{1}{\sqrt{2}}\right) & = & 0.303949\label{Dzz_intersub}\\
\bar{D}_{zz}\left(1,0\right) & = & -0.0197309\label{DzzColsoutside}
\end{eqnarray}
for the nearest neighbors in the other sublattice and the nearest
neighbors in the same sublattice, respectively. The first line above
suggests ferromagnetic ordering of neighboring colums and thus ferromagnetic
ordering of the whole crystal. The total dipolar field at any site
in the depth of the crystal is the sum of the field produced by the
same column and by a few neighboring columns. For ferromagnetic ordering
one finds
\begin{eqnarray}
\bar{D}_{zz}^{(f\mathrm{erro})} & \cong & \bar{D}_{zz}(0,0)+4\bar{D}_{zz}\left(\frac{1}{\sqrt{2}},\frac{1}{\sqrt{2}}\right)+4\bar{D}_{zz}\left(1,0\right)\nonumber \\
 & = & 10.5343\label{DzzFerro}
\end{eqnarray}
that coincides with the value 10.53 previously found by a direct summation
of dipolar field contributions \cite{garchu08prb}. In the case of
alternating ferromagnetic columns within each sublattice one obtains
\begin{equation}
\bar{D}_{zz}^{(110)}\cong\bar{D}_{zz}(0,0)-4\bar{D}_{zz}\left(1,0\right)=9.47634\label{Dzz110}
\end{equation}
 that is very close to the value 9.475 found directly \cite{garchu08prb}.

Let us consider now the effect of the surface. One can calculate the
reduced dipolar field produced by a semi-infinite ordered column of
spins at a site at a distance $\rho=an_{\rho}\equiv a\sqrt{n_{a}^{2}+n_{b}^{2}}$
away from the column and the vertical distance $z=cn_{c}$ from the
end of the column (the surface of the crystal). It is given by
\begin{equation}
\bar{D}_{zz}(\rho,z)=\eta\sum_{n=-n_{c}}^{\infty}\frac{2\eta^{2}n^{2}-n_{\rho}^{2}}{\left(n_{\rho}^{2}+\eta^{2}n^{2}\right)^{5/2}}.
\end{equation}
 For $\rho\gg a$, i.e., $n_{\rho}\gg1,$ one can replace summation
by integration and obtain
\begin{equation}
\bar{D}_{zz}(\rho,z)=a^{2}\int_{-z}^{\infty}dz'\frac{2z'^{2}-\rho^{2}}{\left(\rho^{2}+z'^{2}\right)^{5/2}}=-\frac{a^{2}z}{\left(\rho^{2}+z{}^{2}\right)^{3/2}}.
\end{equation}
One can see that as the observation point moves inside the crystal,
$z\rightarrow\infty,$ the dipolar field disappears. Physically, the
magnetic field produced by a continuous semi-infinite row of magnetic
dipoles only exits at the end of the row and has the same form as
the field produced by a magnetic charge. Writing the result in the
form
\begin{equation}
D_{zz}(\rho,z)=-\frac{a^{2}}{\rho^{2}+z^{2}}\frac{z}{\sqrt{\rho^{2}+z^{2}}}\label{DzzColumnReal}
\end{equation}
one can see that the first fraction is the strength of the reduced
magnetic field $\mathbf{B}^{(D)}$of a point magnetic charge while
the second fraction is the projection cosine of the vector $\mathbf{B}^{(D)}$
upon the axis $c$.

The reduced interaction energy of two semi-infinite ordered columns
at the distance $\rho\gg a$ from each other is given by the integral
\begin{eqnarray}
e(\rho) & = & -\int_{0}^{\infty}\frac{dz}{c}\bar{D}_{zz}(\rho,z)\nonumber \\
 & = & \frac{a^{2}}{c}\int_{0}^{\infty}dz\frac{z}{\left(\rho^{2}+z{}^{2}\right)^{3/2}}=\frac{a^{2}}{c\rho}.
\end{eqnarray}
 The interaction energy in real units $E\left(\rho\right)=E_{D}e(\rho)$
is given by
\begin{equation}
E\left(\rho\right)=\frac{q_{m}^{2}}{\rho},\qquad q_{m}=\frac{Sg\mu_{B}}{c}.\label{ECoulomb}
\end{equation}
This means that two semi-infinite columns interact like two point
changes $q_{m}$ at the ends of the columns.

\begin{figure}
\begin{centering}
\includegraphics[width=8cm]{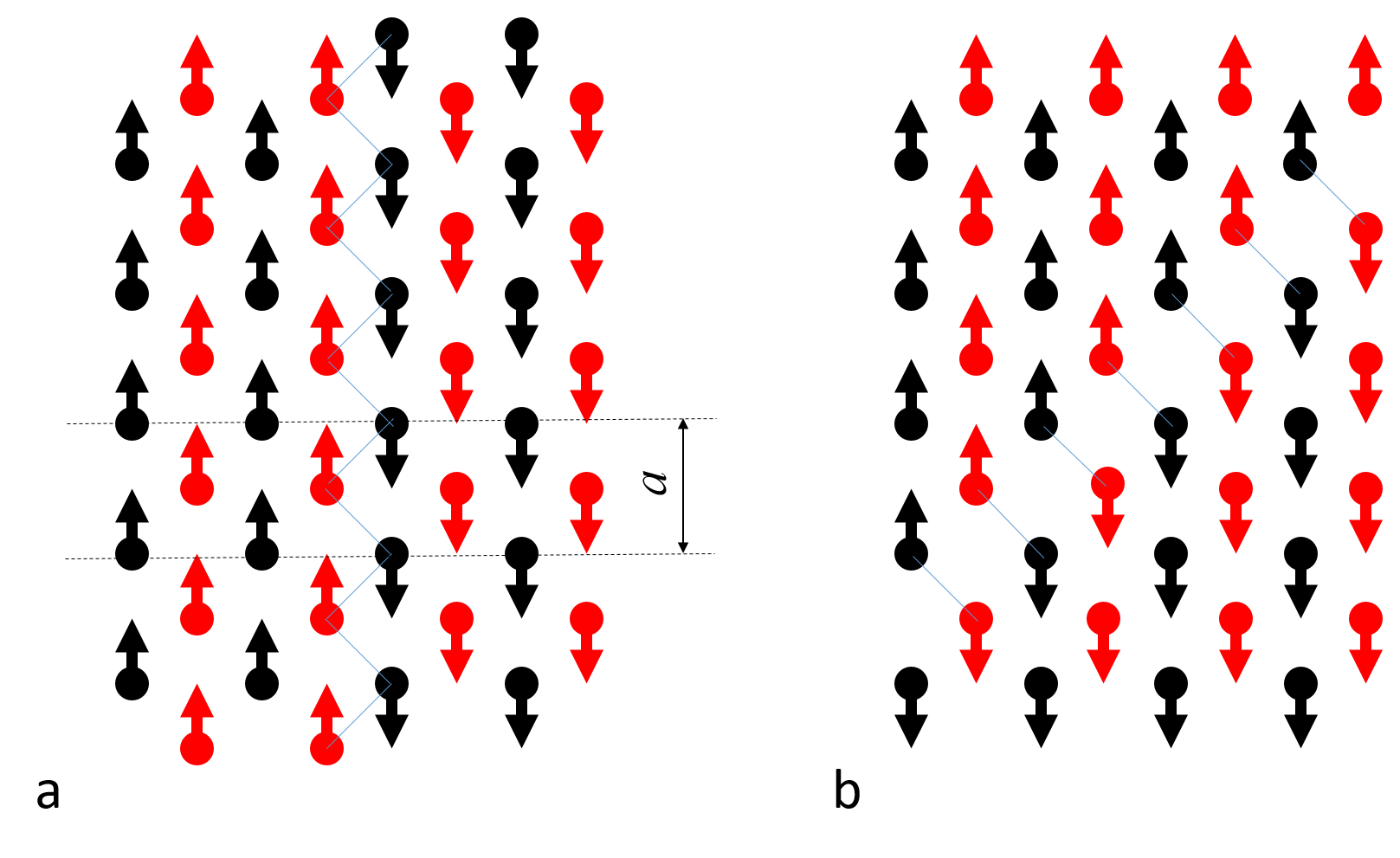}
\par\end{centering}
\caption{Crystal-plane (a) vs diagonal (b) domain walls. The latter have the
energy per unit area by the factor $\sqrt{2}$ lower.\label{Fig_domain_walls}}
\end{figure}

Considering a slab of dimensions $L_{a}$, $L_{b}$, $L_{c}$ and
using the density of the columns $1/(ab)=1/a^{2}$, the total energy
associated with the ferromagnetically ordered surface can be estimated
as
\begin{eqnarray}
E_{\mathrm{surf}}^{(\mathrm{ferro)}} & \sim & \frac{E_{D}}{\left(a^{2}\right)^{2}}\intop\intop dxdydx'dy'e(\rho)\nonumber \\
 & = & \frac{E_{D}}{a^{3}}\intop\intop\frac{dxdydx'dy'}{\rho}\sim E_{D}\frac{L_{\bot}^{3}}{a^{3}},\label{E_surf_ferro}
\end{eqnarray}
where $L_{\bot}\sim L_{a}\sim L_{b}$. This energy is positive that
makes uniform ferromagnetic ordering of columns unfavorable and causes
its splitting into domains. From Fig. \ref{Fig_domain_walls} one
can see that diagonal domain walls (b) have the energy per unit area
by the factor $\sqrt{2}$ lower than domain walls along crystallographic
planes (a). The domain-wall energy is mainly due to the intersublattice
interaction, Eq. (\ref{Dzz_intersub}), and is equal to
\begin{equation}
\frac{1}{2}E_{D}\bar{D}_{zz}\left(\frac{1}{\sqrt{2}},\frac{1}{\sqrt{2}}\right)\frac{2}{c\left(a/\sqrt{2}\right)}=\sqrt{2}\times0.304\frac{E_{D}}{ac}
\end{equation}
per unit area.

To estimate the domain size, one can assume that the crystal is split
into a checkered pattern of up- and down-domains of the size $d$,
so that there are $m=L_{\bot}/d$ domains in each transverse direction.
The surface energy is due to the self-interaction within the islands
of size $d^{2}$ and between the islands. Modification of Eq. (\ref{E_surf_ferro})
yields the estimation
\begin{equation}
E_{\mathrm{surf}}^{(\mathrm{domains)}}\sim m^{2}E_{D}\frac{d^{3}}{a^{3}}=E_{D}\frac{L_{\bot}^{2}d}{a^{3}}\label{ESurfDomains}
\end{equation}
that is smaller than Eq. (\ref{E_surf_ferro}). These domains create
$\sim m$ domain walls having the energy
\begin{equation}
E_{\mathrm{vol}}^{(\mathrm{DWs)}}\sim mE_{D}\frac{L_{\bot}L_{c}}{ac}\sim E_{D}\frac{L_{\bot}^{2}L_{c}}{acd},
\end{equation}
where factors of order 1 have been discarded. Minimizing the total
excessive energy due to the surface charges and domain walls per magnetic
molecule
\begin{equation}
\delta E=\left(E_{\mathrm{surf}}^{(\mathrm{domains)}}+E_{\mathrm{vol}}^{(\mathrm{DWs)}}\right)\frac{a^{2}c}{L_{\bot}^{2}L_{c}}=E_{D}\left(\frac{d}{L_{c}}+\frac{a}{d}\right)
\end{equation}
 with respect to $d$, one obtains
\begin{equation}
d\sim\sqrt{aL_{c}},\qquad\delta E\sim E_{D}\sqrt{\frac{a}{L_{c}}}.\label{eq:d_and_delta_E}
\end{equation}
 As for ordinary ferromagnets, the period of the domain structure
$d$ does not depend on the transverse size of the crystal and increases
with crystal length. Independence of the transverse size means independence
of the aspect ratio for the box-shape crystals, in contrast to the
crystals of the ellipsoidal shape (see the end of Sec. \ref{sec:The-model}).
This finding will be checked numerically below.

Let us now estimate the magnetic susceptibility in the domain state
assuming that applied magnetic field slightly shifts domain walls
making the sizes of positive and negative domains different: $d\Rightarrow d_{\pm}=d\pm\delta d$.
This changes spin polarization per molecule as
\begin{equation}
\left\langle \sigma_{z}\right\rangle =\frac{d_{+}^{2}-d_{-}^{2}}{d_{+}^{2}+d_{-}^{2}}\cong\frac{2\delta d}{d}.
\end{equation}
The corresponding Zeeman energy is $-E_{D}\left\langle \sigma_{z}\right\rangle \tilde{B},$
where the reduced magnetic field $\tilde{B}$ is defined by $B=\left(Sg\mu_{B}/v_{0}\right)\tilde{B}.$
While the energy of domain walls does not change, the changed surface
energy due to the spin polarization $\left\langle \sigma_{z}\right\rangle $,
averaged over the domain structure, can be estimated by adding the
factor $\left\langle \sigma_{z}\right\rangle ^{2}$ to Eq. (\ref{E_surf_ferro})
that yields the additional energy
\begin{equation}
E_{\mathrm{surf}}^{(B)}\sim E_{D}\frac{L_{\bot}^{3}}{a^{3}}\left\langle \sigma_{z}\right\rangle ^{2}\sim E_{D}\frac{L_{\bot}^{3}}{a^{3}}\frac{(\delta d)^{2}}{d^{2}},
\end{equation}
c.f. Eq. (\ref{ESurfDomains}). Thus the total energy per spin becomes
\begin{equation}
\delta E=E_{D}\left(\frac{d}{L_{c}}+\frac{L_{\bot}}{L_{c}}\frac{(\delta d)^{2}}{d^{2}}+\frac{a}{d}-\frac{\delta d}{d}\tilde{B}\right).
\end{equation}
Minimizing this energy with respect to $\delta d$, one obtains $\delta d\sim\left(L_{c}/L_{\bot}\right)d\tilde{B}$.
The corresponding reduced susceptibility per magnetic molecule becomes
\begin{equation}
\tilde{\chi}=\frac{\left\langle \sigma_{z}\right\rangle }{\tilde{B}}\sim\frac{\delta d}{d\tilde{B}}\sim\frac{L_{c}}{L_{\bot}},\label{chi_Def}
\end{equation}
the aspect ratio. For long crystals domains are strongly coupled to
the magnetic field because of their length, thus the susceptibility
can be very large. Finally, the saturation field can be estimated
as
\begin{equation}
\tilde{B}_{s}\sim\frac{1}{\tilde{\chi}}\sim\frac{L_{\bot}}{L_{c}},\label{eq:B_saturation}
\end{equation}
whereas in real units $B_{s}=\left(Sg\mu_{B}/v_{0}\right)\tilde{B}_{s}$.

\section{Numerical method\label{sec:Numerical-method}}

Equilibrium properties of the molecular magnet Mn$_{12}$Ac were studied
by the Metropolis Monte Carlo method making successive trial spin
flips $\sigma_{iz}\rightleftharpoons-\sigma_{iz}$ at lattice sites
$i$ and computing the ensuing energy change $\Delta E=-E_{D}D_{i,zz}\Delta\sigma_{iz}$.
In the case $\Delta E<0$ the trial is accepted, while for $\Delta E>0$
the trial is accepted with the probability $\exp(-\Delta E/T)$. The
trials themselves were done with the probability of 1/2 only that
ensures a true stochastic behavior. If all trials are done, at high
temperatures nearly 100\% acceptance rate results in a nearly deterministic
process leading to the state of the highest energy.

Early papers using Monte Carlo sumulations of small-size molecular
magnets, e.g., Ref. \cite{feralo00prb}, do not report the details
of the procedure. For crystals of large size, however, the long-range
dipolar interaction makes a straightforward application of the Monte
Carlo procedure very slow, so that it has to be optimized.

First, the dipolar field in the crystal has to be computed by a procedure
based on the fast Fourier transform (FFT) that takes the time nearly
linear in the system size (the number of magnetic molecules $N$),
in contrast to the direct summation in Eq. (\ref{BviaD}) that takes
the time growing as $N^{2}$. Wolfram Mathematica that was used for
computations here, allows to obtain dipolar fields at each site $i$
by the $\mathsf{ListConvolve}$ command making summation in Eq. (\ref{BviaD})
by a procedure internally implementing the FFT.

Second, even using the FFT takes a time $\sim N^{2}$ for a full system
update, if the dipolar fields are recalculated before or after each
trial. To reduce this time, the following approach was taken. Both
spin values $\sigma_{iz}$ and the dipolar fields $D_{i,zz}$ are
being fed into the Monte Carlo routine. In the case of a rejected
trial, dipolar fields are not recalculated. In the case of accepted
trial, only the dipolar fields on the neighboring sites were recalculated,
here on up to the 5 neighboring sites along the $c$-direction and
up to 2 neighboring sites in the perpendicular directions. This procedure
is especially well justified at high temperatures, where spins are
disordered and there is no long-range contribution to the dipolar
field, whereas the short-range contribution decreases fast enough
as $1/r^{3}$. At lower temperatures, the acceptance rate decreases
and the change of the long-range field is small again. In any case,
recalculating the short-range part of the dipolar field captures the
main part of the effect and does not take much time so that the full
system update takes a time $\sim N$. After the full system update
of spins, one can make the global update of the dipolar field using
the FFT-based $\mathsf{ListConvolve}$ command that again takes a
time $\sim N$.

Measurements of the spent computer time showed that global updating
of the dipolar fields, even with the FFT, takes much longer than the
Monte Carlo updating of spins with recomputing the short-range fields.
This allows a firther speed-up at low temperatures where acceptance
rate of spin flips is low and the change of dipolar fields in one
system update is small. One can make the costy global dipolar-field
update only after a certain percentage of spins flipped during a number
of full spin updates. Tests have shown that for large-size systems
this maximal fraction of flipped spins is about 2\%. Increasing this
fraction to 3\% and more leads to breaking columns of spins at the
ends of the crystal at low temperatures that is an artifact of the
approximation made. Indeed, at low temperatures, as the result of
ordering, the long-range dipolar field becomes more important and
less tolerant to approximations.

\begin{figure}
\begin{centering}
\includegraphics[width=8cm]{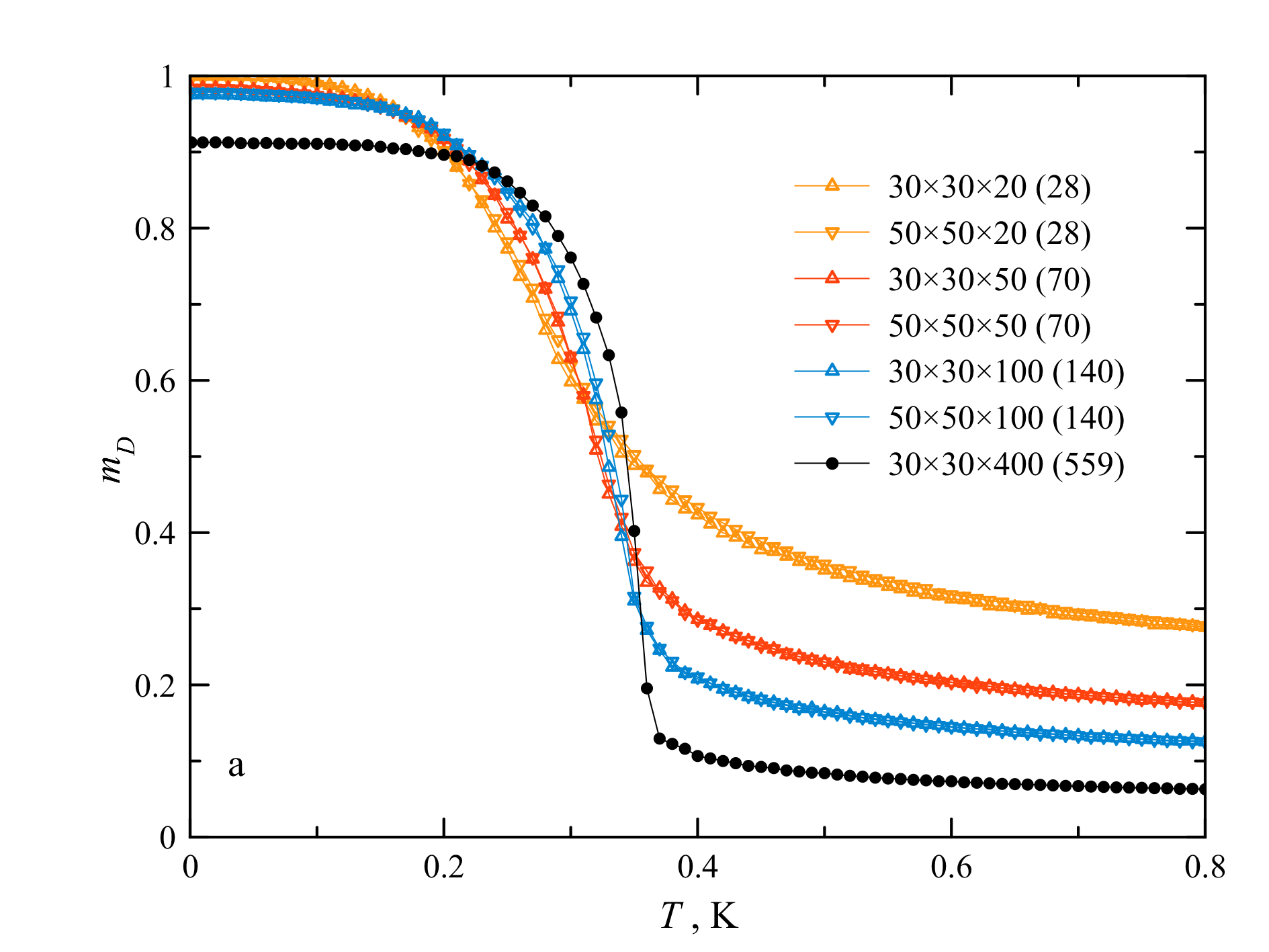}
\par\end{centering}
\begin{centering}
\includegraphics[width=8cm]{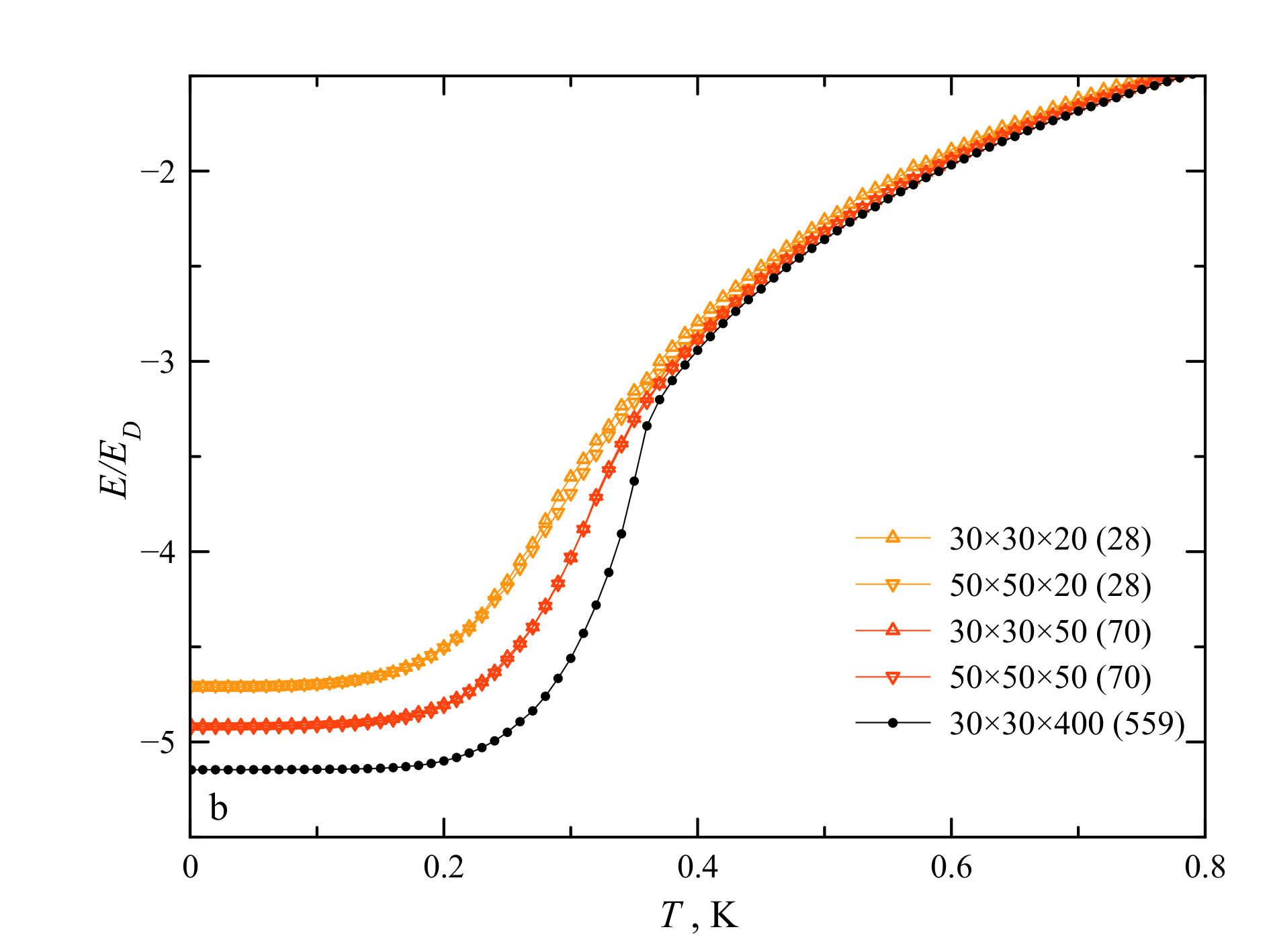}
\par\end{centering}
\caption{Temperature dependence of (a): domain order parameter $m_{D}$ and
(b): reduced energy of box-shape crystals of Mn$_{12}$Ac. The shapes
are labeled by $L_{a}\times L_{b}\times L_{c}(N_{c})$, where the
linear sizes $L$ are in the units of $a$ and $N_{c}=L_{c}/\eta$.
The number of magnetic molecules in the crystal is $2\times L_{a}\times L_{b}\times N_{c}$.\label{Fig_T-dependences}}
\end{figure}

The stopping/measuring strategy used in the applied Monte Carlo procedure
is the following. A block of Monte Carlo spin updates of the size
NMC is defined, in most cases $\mathrm{NMC=100}$. This is the minimal
number of Monte Carlo updates for each temperature or bias magnetic
field. Inside this block of updates, the energy of the system was
monitored and its mean value and dispersion were computed. The energy
change over NMC updates was estimated as the double difference of
the mean energy values over the first and second halfs of the NMC
interval. When the energy change became less than a fraction of the
energy dispersion, here 0.2, it was concluded that the system reached
equilibrium, the procedure was stopped and the averages of physical
quantities over the block of last NMC updates were computed. Above
$T_{C}$ equilibration was very fast, so that the total number of
updates was only slightly higher than NMC. Near $T_{C}$ critical
slowing down was observed, and the total number of updates needed
to reach equilibrium increased by a factor about three. Such quantities
as the energy and magnetization are self-averaging with increasing
the size of the system, so that for the sizes of about one million
magnetic molecules $\mathrm{NMC=100}$ is sufficient to obtain good
smooth data.

\begin{figure}
\begin{centering}
\includegraphics[width=8cm]{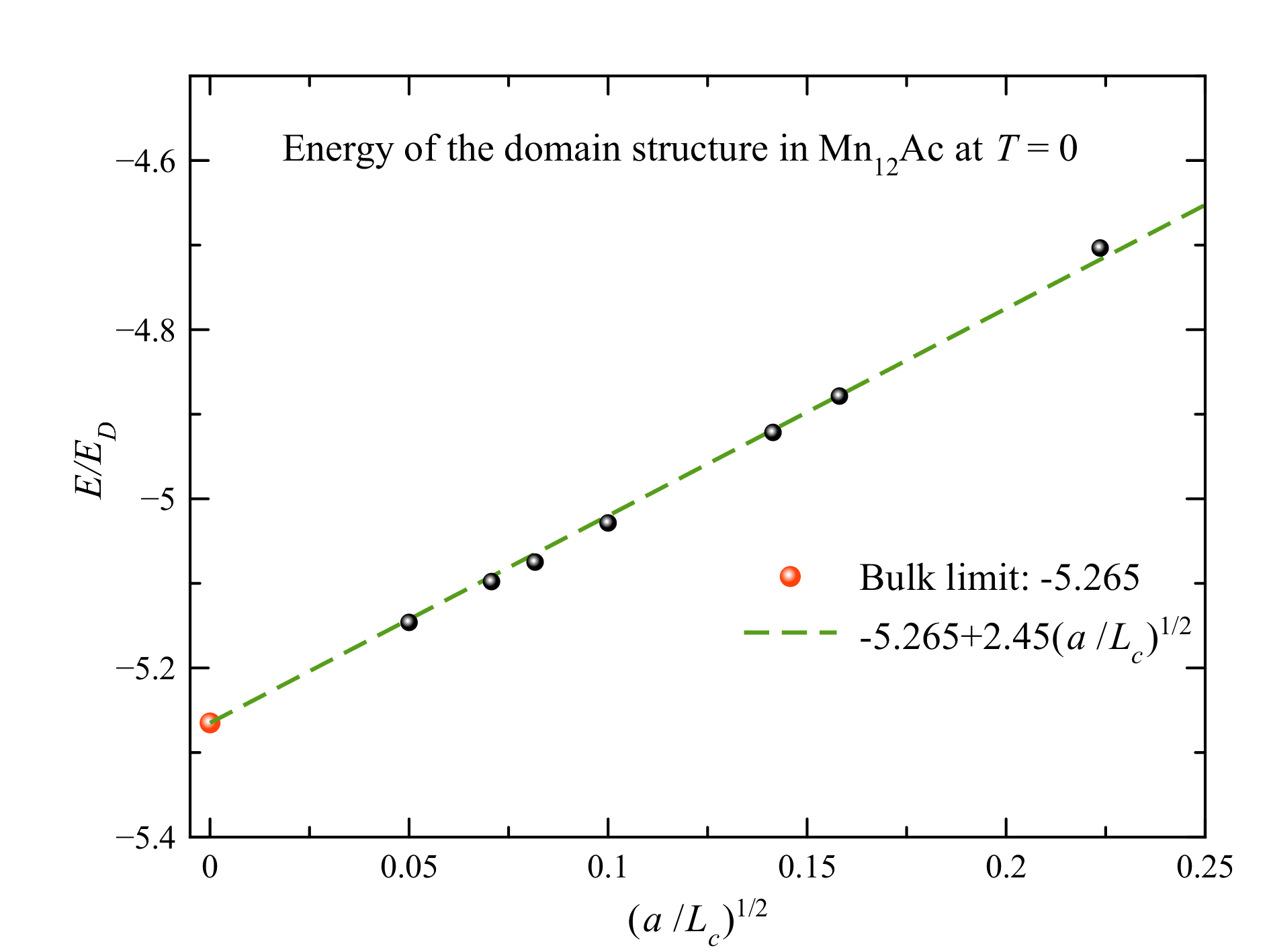}
\par\end{centering}
\caption{Dependence of the energy per magnetic molecule at $T=0$ on crystal
length, showing the contribution of surfaces and domain walls.}

\label{Fig_E_vs_length}
\end{figure}

With the conditional global dipolar-field updating, computation tends
to be very fast at low temperatures because of the low acceptance
rate. On the other hand, for a good system equilibration one needs
a sufficient number of spin-flip acceptances. Thus the stopping/measuring
strategy was modified as follows. The computation at a given temperature
or bias field was stopped only after a certain fraction of acceptances
per spin at high temperatures, $0.5\mathrm{NMC},$ was achieved. In
most cases $0.2\times0.5\mathrm{NMC}$ acceptances per spin was required.
In the computation control program fulfillment of both criteria was
required: equilibration and sufficient number of acceptances. The
averaging interval was defined as the maximum of NMC updates and the
last 1/4 of the actual number of spin updates, the latter becoming
the case at low temperatures, exceeding NMC by far. Finally, the upper
limit on the number of spin updates was set, typically 3000, to avoid
getting stuck at the lowest temperatures.

Because of the global operation of computing the dipolar field, the
computation is not fully parallelizable. Although the spin-update
part can be parallelized, it does not make much sense because it is
much faster than the global dipolar-field update even in its non-parallelized
version.

Since the space-average magnetization in the domain state is zero,
the domain order parameter
\begin{equation}
m_{D}=\sqrt{\left\langle \left(\frac{1}{N_{c}}\sum_{n_{c}=1}^{N_{c}}\sigma_{iz}\right)^{2}\right\rangle _{\mathrm{cols,}\mathrm{subs}}},
\end{equation}
was computed for any state generated in the Monte Carlo routine and
averaged over these states within the averaging interval. For all
columns perfectly ordered, that is the case for a perfect domain state,
one has $m_{D}=1$. For the completely disordered state, one has $m_{D}=1/N_{c}$,
where $N_{c}=L_{c}/c$ is the number of magnetic molecules in the
column.

\begin{figure}
\begin{centering}
\includegraphics[width=7cm]{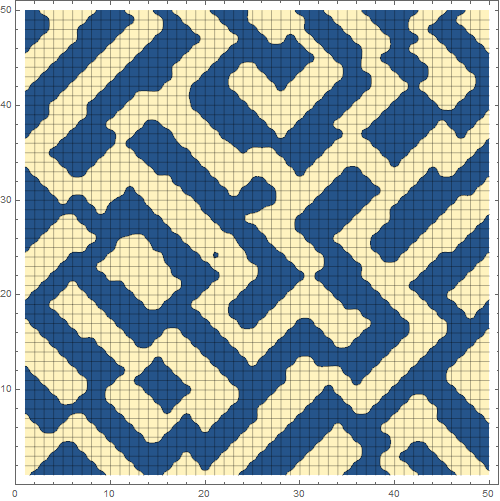}
\par\end{centering}
\caption{Domains in the Mn$_{12}$Ac crystal of size $50\times50\times50$(70)
at $T=0$ obtained by slow cooling, view along the $c$-axis. The
plot was created using average spin polarization in $c$-columns.}

\label{Fig_Domains_ab}
\end{figure}
\begin{figure}
\begin{centering}
\includegraphics[width=8cm]{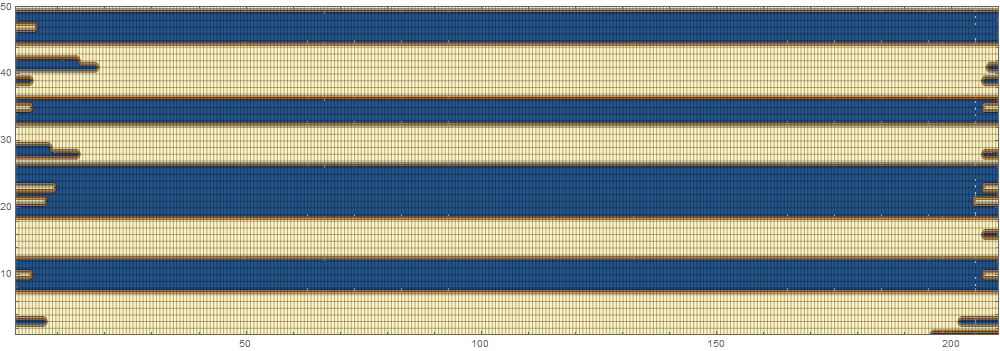}
\par\end{centering}
\caption{Domains in the Mn$_{12}$Ac crystal of size $50\times50\times150$(210)
at $T=0$ obtained by slow cooling, in the central $ac$ cross-section.\label{Fig_Domains_ac}}
\end{figure}

The reduced linear susceptibility $\tilde{\chi}$ can be computed
either using the definition in Eq. (\ref{chi_Def}) or using the formula
\begin{equation}
\tilde{\chi}=\frac{\left\langle m^{2}\right\rangle -\left\langle m\right\rangle ^{2}}{\tilde{T}},\quad m\equiv\frac{1}{N}\sum_{i}\sigma_{iz},\quad\tilde{T}\equiv\frac{T}{E_{D}}.\label{chi_fluct}
\end{equation}
The latter is not self-averaging for large system sizes and requires
very long Monte Carlo sequences to average out fluctuations that is
too costy for large systems.

\section{Numerical results\label{sec:Numerical-results}}

Results for the temperature dependence of the domain order parameter
$m_{D}$ and the energy of box-shape Mn$_{12}$Ac crystals of different
sizes obtained by lowering the temperature in small steps are shown
in Fig. \ref{Fig_T-dependences}. For the largest crystal containing
slightly above $10^{6}$ magnetic molecules, the computation requires
about 30 hours. In accordance with the conclusion of the preceding
section, the results depend only on the length of the crystal but
not on its transverse dimensions. In the limit $T\rightarrow0$ there
are deviations of $m_{D}$ from the theoretical value 1 since column
shapes do not fully equilibrate, especially for the longest crystal,
so that domain walls are not everywhere parallel to the $c$-axis.
Here annealing could help but it was not done.

The Curie temperature for the longest crystal is $T_{C}=0.36$ K,
about two times lower than its mean-field value $0.707$ K. Such a
big difference speaks of large critical fluctuations that can be expected
in this quasi-one-dimensional systems with much stronger interaction
within $c$-columns. The computed value of $T_{C}$ is very close
to the experimentally measured value $T_{C}=0.34$ K for Fe$_{8}$
in Ref. \cite{burzurietal11prl}. Since the two materials have comparable
lattice parameters and the mean-field Curie temperature in Fe$_{8}$
is about 1 K, the close values of $T_{C}$ for the two materials are
not surprizing. Fe$_{8}$ has even smaller ratio $T_{C}/T_{C}^{\mathrm{MFA}}$
that could be tentatively explained by the pyramidal shape of the
crystals less favorable for ferromagnetic ordering. Unfortunately,
this shape makes computation of the dipolar field in Fe$_{8}$ difficult.

\begin{figure}
\begin{centering}
\includegraphics[width=8cm]{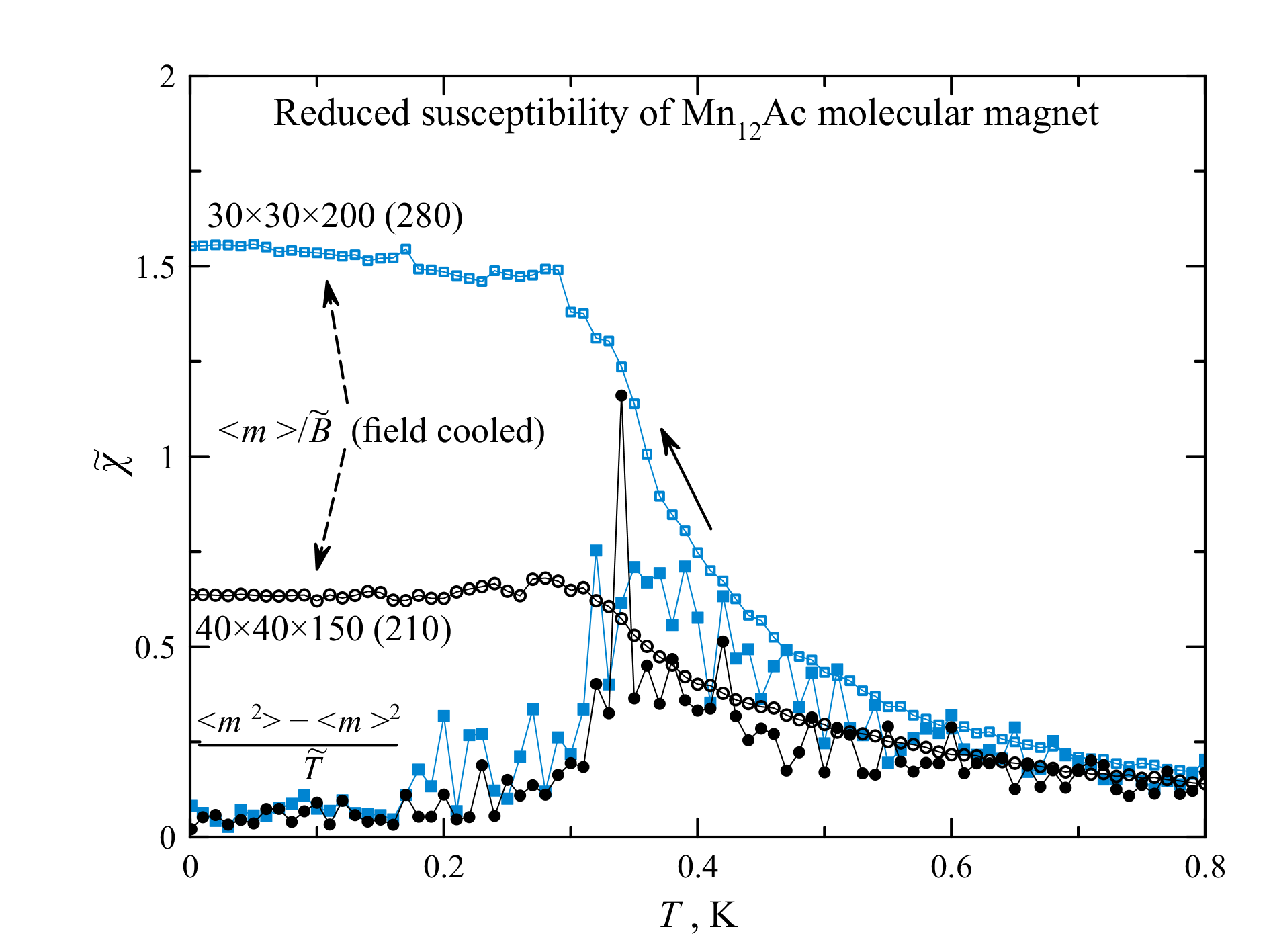}
\par\end{centering}
\caption{Reduced susceptibility of Mn$_{12}$Ac computed by two methods shows
that domain walls are pinned at low temperatures.\label{Fig_chi_vs_T}}
\end{figure}

\begin{figure}
\begin{centering}
\includegraphics[width=8cm]{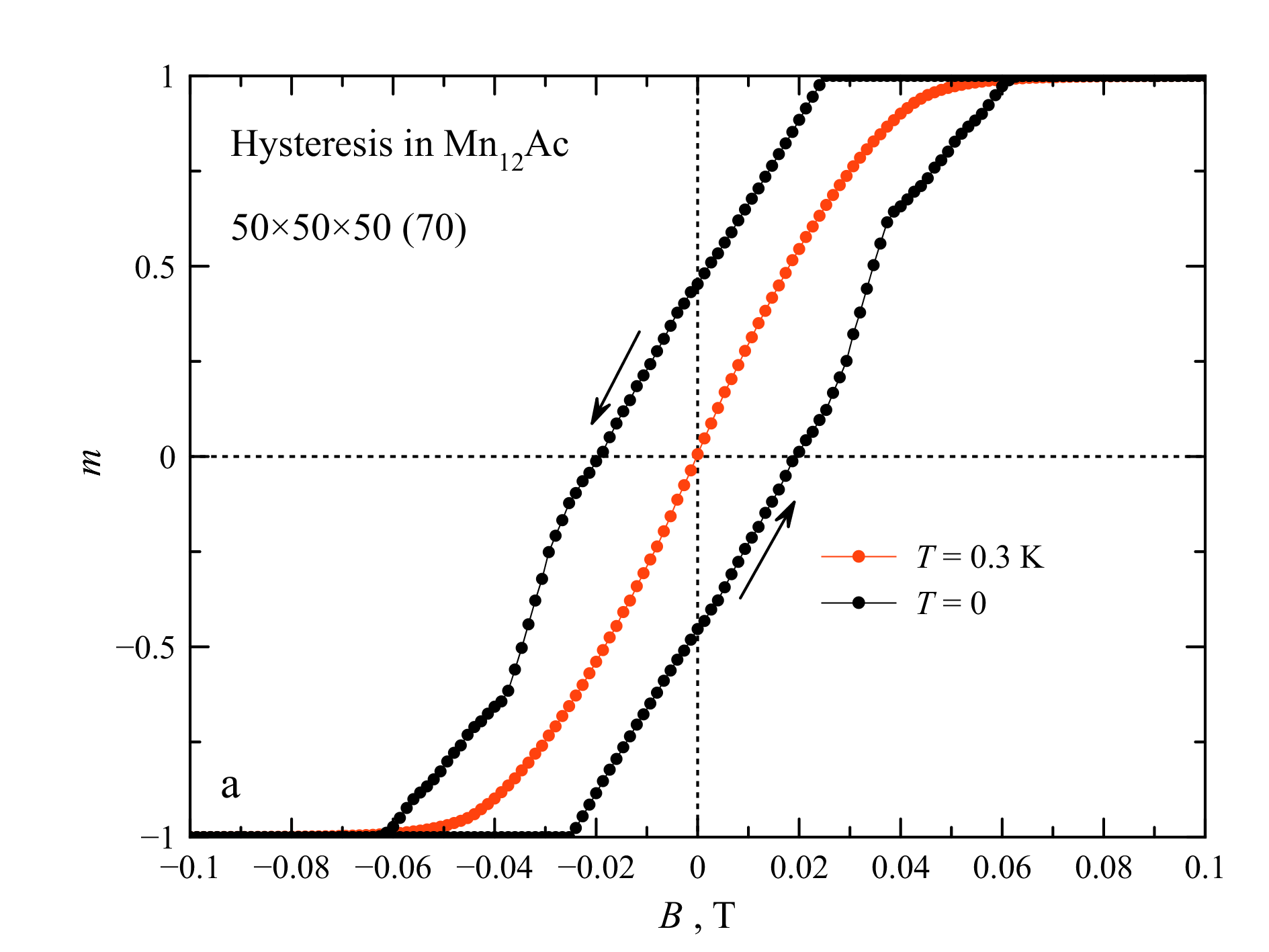}
\par\end{centering}
\begin{centering}
\includegraphics[width=8cm]{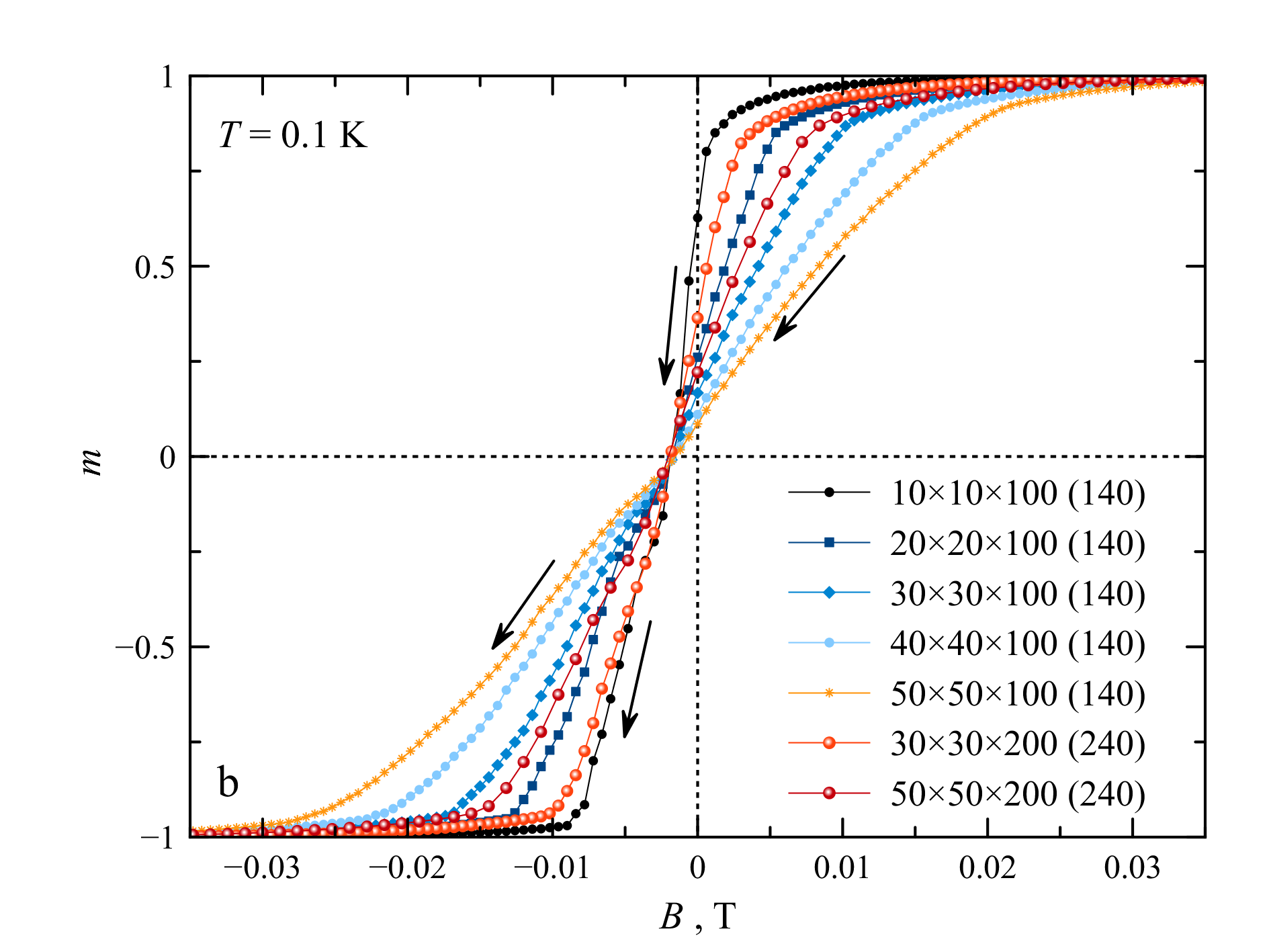}
\par\end{centering}

\caption{Magnetization curves in Mn$_{12}$Ac (a): Hysteresis loop is vanishing
with increasing temperature. (b) Equilibrium magnetization curves
of Mn$_{12}$Ac with different transverse size of the crystal. }

\label{Fig_m_vs_H}
\end{figure}

Fig. \ref{Fig_E_vs_length} confirms Eq. (\ref{eq:d_and_delta_E})
for the contribution of the surface and domain walls to the energy
and Eq. (\ref{E0Def}) in the bulk limit. By fitting one obtains $\delta E/E_{D}\simeq2.45\sqrt{a/L_{c}}$.
Since crystals used in the experiments are macroscopic, domains are
large and the contribution of surfaces and walls into the energy is
small.

Fig. \ref{Fig_Domains_ab} obtained by slow cooling the sample by
step-wise lowering the temperature shows that domain walls are indeed
diagonal, as suggested in Sec. \ref{sec:Columns-and-domains}. For
plotting, the average spin polarization in each column was used.

In Fig. \ref{Fig_Domains_ac} that provides a view onto the $ac$
cross-section with $n_{b}=N_{b}/2$, one can see that domains are
parallel to the $c$-axis. At the ends of the crystal there is a little
disturbance of the main structure that can be interpreted as domain
branching.

The pattern of the dipolar field $D_{i,zz}$ is the same as that of
$\sigma_{iz}$, so that figures such as Fig. \ref{Fig_Domains_ab}
for these two quantities are indistinguishable. Domain walls are atomic-narrow,
and on both sides of the wall $D_{i,zz}$ has its almost maximal positive
or negative value created by the column of up or down spins. Since
each spin, even adjacent to the wall, is polarized by a strong field,
Monte Carlo spin flips are almost 100\% rejected at low temperatures.
This means that domain walls are pinned by the lattice and cannot
easily move.

Lack of thermodynamic equilibrium at low temperatures is strikingly
manifested in Fig. \ref{Fig_chi_vs_T} showing reduced susceptibility
computed by two different methods. Slow cooling in a small fixed magnetic
field yields $\tilde{\chi}=\left\langle \sigma_{z}\right\rangle /\tilde{B}$
that increases with decreasing temperature and reaches a plateau.
In computations, the value $\tilde{B}=0.1$ was used, so that the
low-temperature magnetization $m=\tilde{\chi}\tilde{B}\approx0.15$
is still far from saturation and the susceptibility should be close
to linear. Computation with $\tilde{B}=0.03$ yields similar results
with more scatter (not shown). From the plateau values one obtains
$\tilde{\chi}L_{\bot}/L_{c}=0.23$ for the $30\times30\times200$
crystal and 0.17 for the $40\times40\times150$ crystal. These values
qualitatively support Eq. (\ref{chi_Def}). The low-temperature susceptibility
could be interpreted as due to shiftifting of domain walls by the
applied field, as in common ferromagnets. However, actual changing
the field from zero to this value would result in a smaller and poorly
defined value of $\tilde{\chi}$ because of domain-wall pinning and
Barkhausen jumps. The true linear susceptibility is given by Eq. (\ref{chi_fluct})
and this one becomes small at low temperatures, that can be seen in
Fig. \ref{Fig_chi_vs_T} in spite of rather strong fluctuations that
are very difficult to average out.

Another manifestation of domain-wall pinning is the hysteresis loop
at $T=0$ seen in Fig. \ref{Fig_m_vs_H}a. With increasing the temperature,
this loop disappears. There is no loop at $T=0.3$ K and a narrow
loop at $T=0.1$ K (not shown). Magnetization curves in Fig. \ref{Fig_m_vs_H}b
for crystals with different transverse sizes at $T=0.1$ K, still
showing some hysteresis (only one branch shown), clearly scale with
the transverse size in the horizontal direction, confirming Eqs. (\ref{chi_Def})
and (\ref{eq:B_saturation}). Scaling and fitting magnetization curves,
one obtains the numerical factor 0.12 in Eq. (\ref{chi_Def}) that
is of the same order of magnitude as found above.

\section{Discussion}

By slow cooling within the Monte Carlo method it was shown that Mn$_{12}$Ac
orders ferromagnetically at 0.36 K into the ferromagnetic state with
domains, extending findings of Ref. \cite{gar10prbrc} where such
type of ordering was obtained for much smaller crystals within the
space-resolved mean-field approximation. Such a low value of $T_{C}$,
two times lower than its mean-field value, is the consequence of the
quasi-one-dimensional ordering dominated by columns of magnetic molecules
along the $c$ axis. It was demonstrated that domains, being parallel
to the $c$-axis, are separated by atomic-narrow domain walls whose
cross-sections consist of diagonal lines in $ab$ planes, Fig. \ref{Fig_Domains_ab}.
Such narrow domain walls are pinned by the lattice at low temperatures.

It was shown that the period of the domain structure and the introduced
here \textit{domain order parameter} $m_{D}$ depend on the length
of the crystal (in the $c$ direction) and do not depend on its transverse
size. Magnetic susceptibility at low temperatures depends on crystal's
aspect ratio, Eq. (\ref{chi_Def}), being large for prolate crystals.

In real Mn$_{12}$Ac spin transitions between up and down states are
very slow at low temperatures, since they require spin tunneling under
the energy barrier. Because of this, the crystal will not order spontaneously.
Spin tunneling can be sped up and ordering can be facilitated if a
strong magnetic field transverse to the anisotropy axes is applied.
However, this can produce the undesirable bias because of the small
misalignment of the anisotropy axes in different Mn$_{12}$Ac crystals.
In this work, the effects of the energy barriers were not taken into
account and a simple Ising-type model was adopted. Another molecular
magnet Fe$_{8}$ is a better candidate for dipolar ordering because
of the faster tunneling via transverse anisotropy, and the ordering
was indeed observed by magnetic measurements at $T_{C}=0.34$ K, close
to the value for Mn$_{12}$Ac found here. However, shapes of Fe$_{8}$
crystals are complicated and computation of dipolar fields is a more
difficult problem.


\end{document}